\documentclass{article}
\usepackage{arxiv}
\usepackage[utf8]{inputenc} 
\usepackage[T1]{fontenc}   
\usepackage[hidelinks]{hyperref}
\usepackage{url}            
\usepackage{booktabs}       
\usepackage{amsfonts}       
\usepackage{nicefrac}       
\usepackage{microtype}      
\usepackage{lipsum}
\usepackage{graphicx}
\usepackage{multirow}
\usepackage{siunitx}
\usepackage{float}

\graphicspath{ {./images/} }

\title{What can we learn from marketing skills as a bipartite network from accredited programs?}

\author{
 Silvana Dakduk\href{https://orcid.org/0000-0003-4700-5354}{\includegraphics[width=0.3cm]{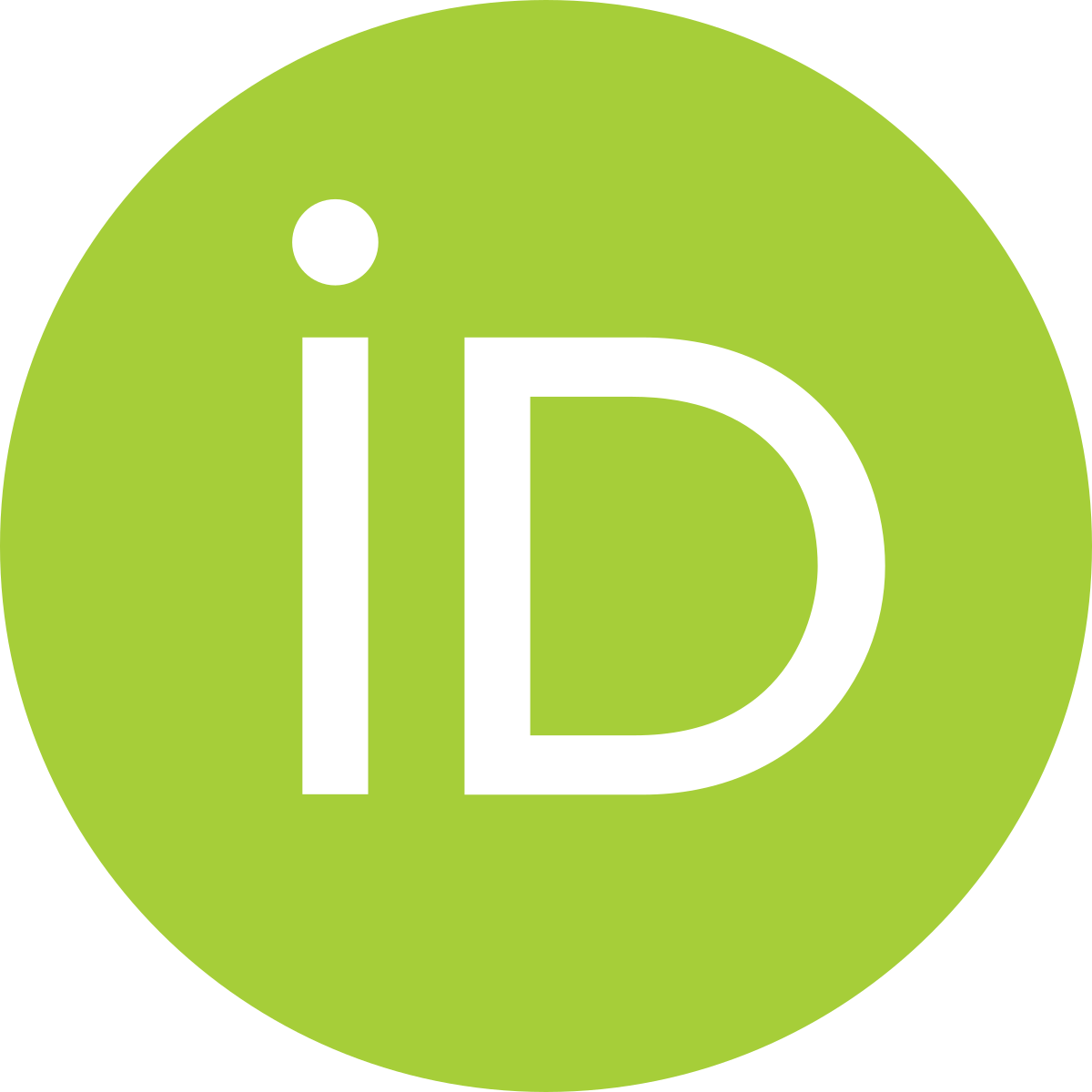}} \\
  School of Management\\
  Universidad de Los Andes\\
  Bogotá, Colombia \\
  \texttt{sm.dakduk@uniandes.edu.co} \\
   \And
 Maria del Pilar García-Chitiva\href{https://orcid.org/0000-0001-6776-3422}{\includegraphics[width=0.3cm]{orcid.png}} \\
  Institute for the Future of Education\\
  Tecnologico de Monterrey\\
  Monterrey, Nuevo Leon, Mexico \\
  \texttt{pilargarciach@tec.mx} \\
  \And
 Juan C. Correa\href{https://orcid.org/0000-0002-0301-5641}{\includegraphics[width=0.3cm]{orcid.png}} \\
  Research \& Development Unit\\
  Critical Centrality Institute\\
  Monterrey, Nuevo Leon, Mexico \\
  \texttt{j.correa.n@gmail.com} \\
}

\begin{document}
\maketitle
\begin{abstract}
The relationship between professional skills and higher education programs is modeled as a non-directed bipartite network with binary entries representing the links between 28 skills (as captured by the occupational information network, O*NET) and 258 graduate program summaries (as captured by commercial brochures of graduate programs in marketing with accreditation standards of the ``Association to Advance Collegiate Schools of Business''). While descriptive analysis for skills suggests a qualitative lack of alignment between the job demands captured by O*NET, inferential analyses based on exponential random graph model estimates show that skills' popularity and homophily coexist with a systematic yet weak alignment to job demands for marketing managers.
\end{abstract}


\section{Introduction}

A challenging decision for any manager is placing the right person in the right job. In search for predictors of job performance, the American psychologist David C. McClelland \cite{Mcclelland1973} argued that academic aptitude, knowledge from content tests, school grades, and credentials are often poor predictors of employee performance. McClelland's research laid the groundwork for the ``job competence assessment'' method, which became foundational to management literature in the latter half of the 20$^{th}$ century \cite{Spencer1993}. Contemporary data-driven research supports McClelland's findings, suggesting that academic offerings often fall short of meeting companies' job requirements \cite{Borner2018}. In this context, gaining a competitive advantage from a labor perspective increasingly relies on improving the educational-occupational match \cite{Gadar2018}. Some scholars argue that one way to enhance this education-job alignment is by estimating the value of employees' skills, especially given the unpredictable future \cite{Vista2020}. 

From a social network perspective, a network researcher can consider that skills represent a set with connections to university programs that represent the academic offering \cite{Garciachitiva2024}. The formalization of these connections is a bipartite network structure represented as a triple $Y = (R,C,E)$. In this triple, $R$ and $C$ refer to two disjoint sets of nodes, with $R$ being the first partition of nodes (i.e., rows in a bi-adjacency matrix) and $C$ being the second partition of nodes (i.e., columns in a bi-adjacency matrix), and $E$ $\subseteq R \times C$ is the set of edges of the network \cite{Kevork2022}. As per \cite{Garciachitiva2024}, this bipartite network facilitates a deeper understanding of skills' centrality in the landscape of the academic offering, which has conceptual roots with limited data-driven evidence. 

A conceptual root of the above-mentioned bipartite network goes back to the ``occupational information network'' (O*NET). According to \cite{Peterson2001}, the goal of O*NET was to ``\textit{promote the effective education, training, counseling, and employment of the American workforce. It should accomplish its purpose by providing a database system that identifies, defines, classifies, and describes economic occupations in an accessible and flexible manner}'' (p. 455).

The data-driven evidence supporting the use of bipartite networks as a tool for modeling the relationship between skills and university programs is more recent than the conceptual roots just mentioned. The empirical work of \cite{Gadar2018} and \cite{Garciachitiva2024} provide relevant pillars to the present work. Apart from these two references, the work of \cite{Conaldi2013} deserves special mention for their initial insights into the modeling perspective. Let us review the goal of these previous works to argue why the present work offers an extension of their findings.

\cite{Conaldi2013} analyzed a case study on open-source software development by reconstructing the bipartite network structure generated by the association between 72 professionals (software developers) and 737 problems (software bugs) they had to solve during an entire development cycle of the free/open source software project Epiphany. Among their results, they found that problem solvers tend to distribute their activity over multiple software bugs. However, these bugs tend not to share multiple contributors, and this dual trend toward de-specialization and exclusivity relied on specific local network dependencies estimated by exponential random graph models. Although \cite{Conaldi2013} did not consider university programs in their case study, their inferential approach based on exponential random graph allowed them to go beyond network-based descriptive statistics and contrast some hypotheses about the local network properties as predictors of the entire network.

\cite{Gadar2018} proposed a bipartite network structure to analyze the educational-occupation match of 15,253 subjects who worked in 402 occupations and graduated from 398 education programs delivered by 52 institutions in Hungary. Their results showed that computer science, nursing, and engineering programs connected to occupations with a scarce source of workers. They also found five components or modules in their network, but the programs-occupation connectivity was stronger in three of these modules. The first module grouped teaching, humanity, and art programs (Germanistic studies), and half of their occupations did not require a higher education diploma, although a quarter of graduates from Germanistic studies worked in the manufacturing and IT sector due in part to their advanced knowledge of the German language. The second module showed connections between business, economics, finance, human resources, social work, nursing, and medical programs with health care, teaching, child care workers, medical technicians, and personal care jobs. The third module revealed the links between engineering, IT, physics, ecology, and earth sciences with production, manufacturing, information managers, life science, and engineering professionals.

\cite{Garciachitiva2024} proposed a bipartite network structure to estimate the importance of soft skills centrality by sampling 230 programs for three types of graduate studies (i.e., Specialization, Master, and Doctorate) with two quality accreditation standards offered by 49 universities in Colombia. Using text mining techniques focused on keyword-in-context search, they identified 31 soft skills (i.e., skills related to socio-emotional behavioral control) in programs' intended learning outcomes. Their results showed that skills' centrality varied as a function of the graduate program without revealing statistically significant differences. In addition, while most central skills tend to be those related to creativity, leadership, and analytical thinking, less central skills were those related to empathy, ethical thinking, and critical thinking. Based on these results, the authors claimed to what extent emphasizing the most visible skills might imply an imbalance in the opportunities to enhance other soft skills, such as ethical thinking.

Compared with existing research, this work extends the evidence on the relationship between professional skills and university programs modeled as a bipartite network. Our contribution elaborates upon the case of marketing programs as they represent a niche of accredited business schools' academic offerings. The rationale behind this case is twofold. On the one hand, business school leaders have made crystal clear why radical innovations are essential for their educational practices, given the potential transformation associated with artificial intelligence and other emerging technologies  \cite{Lorange2019,Schlegelmilch2020}. On the other hand, as the interaction between soft skills and training programs is vital to enhancing employee job performance \cite{Marin2022}, an updated perspective on professional training should consider that artificial intelligence technologies have entered into the classroom \cite{Extance2023}. These technologies can serve as the building blocks for future educational innovations based on skills-programs networks.

\section{Empirical motivation and Theoretical background}

Current teaching practices at universities face several challenges for professors and students. The first challenge concerns the increasing number of published articles and how professionals learn to manage the flood of technical information \cite{Bauer2009}. The second challenge concerns how professionals can follow simple rules for developing good reading habits during graduate school and beyond to stay updated on the advances of their disciplines  \cite{Mendez2018}. Recent evidence shows that cognitive patience is essential to preventing the ``\textit{too long, didn't read}'' effect. This patience is the skill to read with focused and sustained attention and delayed gratification, keeping the discipline to refrain from multitasking or skimming over parts of the text \cite{vandeVen2023}. As information consumers, professionals can struggle to distinguish opinions from facts when reading online material \cite{Brossard2013}, including the texts generated by virtual assistants such as ChatGPT \cite{Extance2023}. As these challenges are equally present in all professional disciplines, it is vital to understand which skills are fundamental for students to progress in their studies. Such an understanding does not discard attempts adapted to each discipline but demands a data-driven endeavor. 

We propose a bipartite network that maps the connections between skills and programs as a formal data-driven endeavor to spot the opportunities to enhance university education. This perspective relies on the marketing of schools and universities \cite{Mcdonald2019}, which is an additional resource for universities that compete in the marketplace for talented students and faculty by investing in faculty and student recruitment to maintain or increase their reputation \cite{Volkwein2006}. Besides reputation, universities can compete with pricing strategies and online delivery systems based on virtual education programs. However, these strategies only work when universities advertise their programs and provide written and multimedia material, such as informational pieces (i.e., brochures), to influence future students' decision-making. Here, leveraging brochures as data input for analytical purposes can provide promising research opportunities \cite{Garciachitiva2024}.

\section{Empirical setting and data}

\subsection{Setting}

Our case elaborates upon the marketing education system \cite{Anderson2023}. We focus on the Master's in Marketing segment from higher education institutions currently accredited by the Association to Advance Collegiate Schools of Business (AACSB). The AACSB accreditation is a globally recognized standard of excellence in business education, ensuring that the programs meet high-quality standards and are constantly updated to reflect the evolving needs of the business world. This system's fundamental unit of analysis is the link between the theoretical understanding of skills as job requirements captured by O*NET \cite{Peterson2001} and the commercial offering of graduate programs \cite{Garciachitiva2024}.

O*NET (Occupational Information Network) is a comprehensive and continuously updated database that provides detailed information about various occupations' characteristics and skill requirements in the United States. Developed and maintained by the National Center for O*NET Development, O*NET is a critical resource for employers, HR professionals, educators, and workers aiming to understand labor market demands. For instance, employers can use O*NET to create job descriptions, HR professionals can design training programs, educators can develop curriculum, and workers can use it to identify the skills needed for career advancement. O*NET measures various aspects of occupations, including tasks, work activities, and worker requirements, focusing on skills, knowledge, and abilities. For this study, we specifically focus on the Worker Requirements: Skills component, which highlights the essential skills needed for successful job performance in specific roles.

The O*NET skills list captures the skills in O*NET's Worker Requirements, rated on a scale from 0 to 100, with higher scores indicating greater importance for the occupation in question \cite{Handel2016}. These values are derived from comprehensive surveys administered to job incumbents and occupational analysts. The collected data is processed through a rigorous algorithm that normalizes and weights the responses, establishing a hierarchy of importance for each skill. This systematic ranking, which ensures that the most critical skills for occupational success are accurately identified and prioritized, provides a reliable framework for evaluating occupational requirements.

The process of data collection in O*NET involves several specific steps. Direct observation is conducted where professionals are observed in their work environment to identify their tasks. Surveys are administered to workers, supervisors, and experts in each field to understand better the functions and skills required. Additionally, document analysis is performed by reviewing job manuals, position descriptions, and other relevant documents to complement and verify the gathered information. The collected data is then processed in order to normalize and weight the responses, ensuring a systematic ranking of the importance of each skill. This rigorous process ensures the reliability and validity of the O*NET data \cite{Autor2023,ONET2022}.

\begin{figure}
\centering
\includegraphics[width=.7\columnwidth]{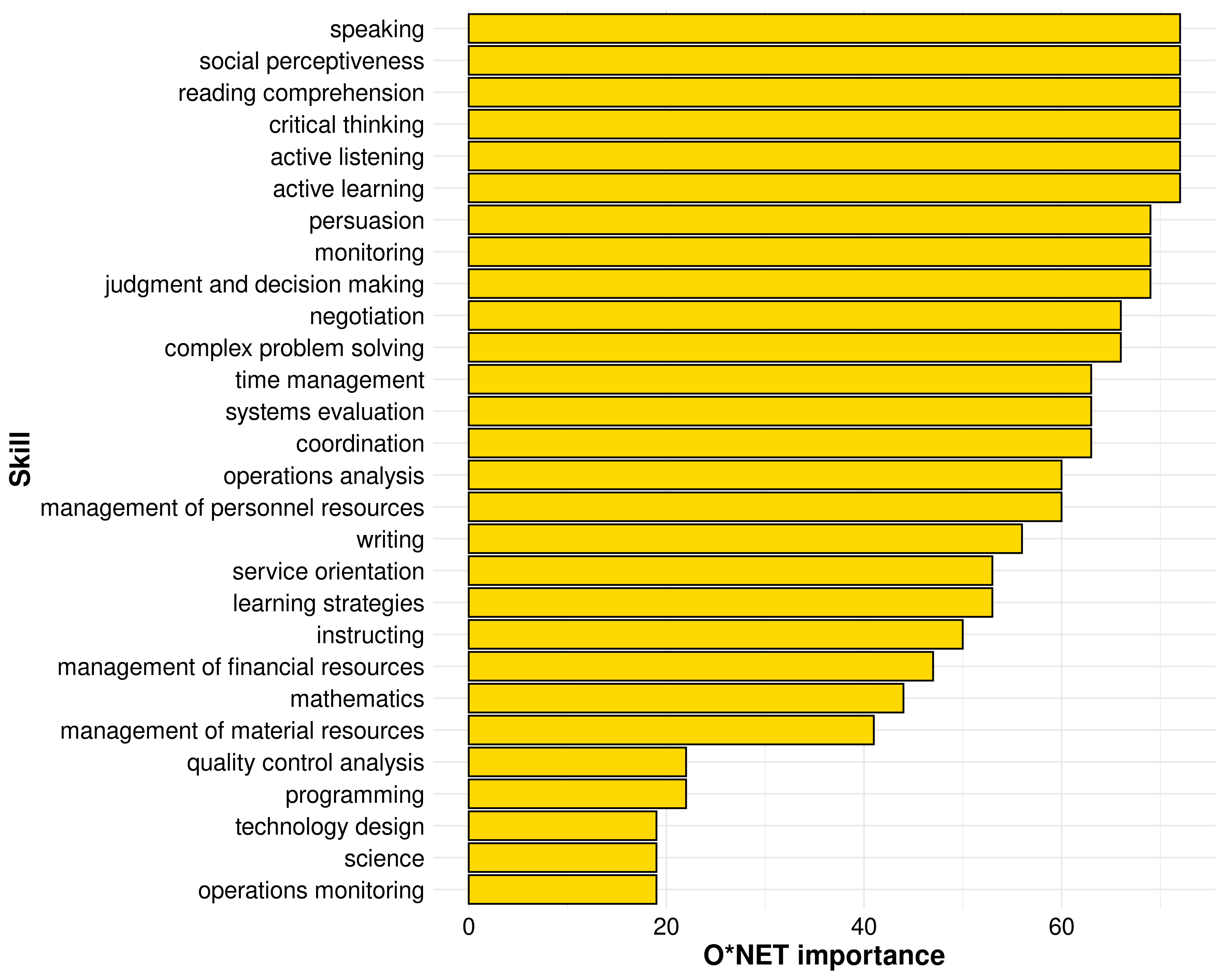}
\caption{Skills' Importance reported by the O*NET profile of Marketing Managers}
\label{FA}
\end{figure}

In this study, we specifically examine the skills identified by O*NET for the role of Marketing Manager (11-2021.00). Marketing Managers are responsible for planning, directing, or coordinating marketing policies and programs, determining the demand for products and services, and developing strategies to maximize market share and profitability. Key job titles within this role include Account Supervisor, Brand Manager, and Marketing Director. Figure \ref{FA} shows the O*NET importance ratings for skills required for the position of Marketing Manager. This representation highlights that among the highest-ranked skills critical for success in this role are soft skills such as ``speaking,'' ``social perceptiveness,'' and ``reading comprehension.''

Our analysis aims to assess the alignment between this hierarchy of skills prioritized by O*NET and the promises made by AACSB-accredited Master's programs in Marketing, as reflected in their promotional brochures. Thus, the extent to which these programs are designed to develop the skills that O*NET identifies as essential for success in marketing management is a proxy to the degree of coherence between the educational offerings and the current demands of the marketing profession.

The AACSB accreditation is a globally recognized standard of excellence in business education, ensuring that the programs meet high-quality standards and are constantly updated to reflect the evolving needs of the business world. As a professional association founded in 1916, AACSB continues to promote a united movement to improve the quality of business education worldwide. This mission is achieved by connecting, sharing, and inspiring innovation and quality throughout its member network and the broader business community. AACSB's collective strength is built on diverse perspectives, a global mindset, and a commitment to making a difference in business education. Members of AACSB share this mindset, particularly in marketing education, which is critical for developing future leaders in the industry.

\begin{figure}
\centering
\includegraphics[width=.9\columnwidth]{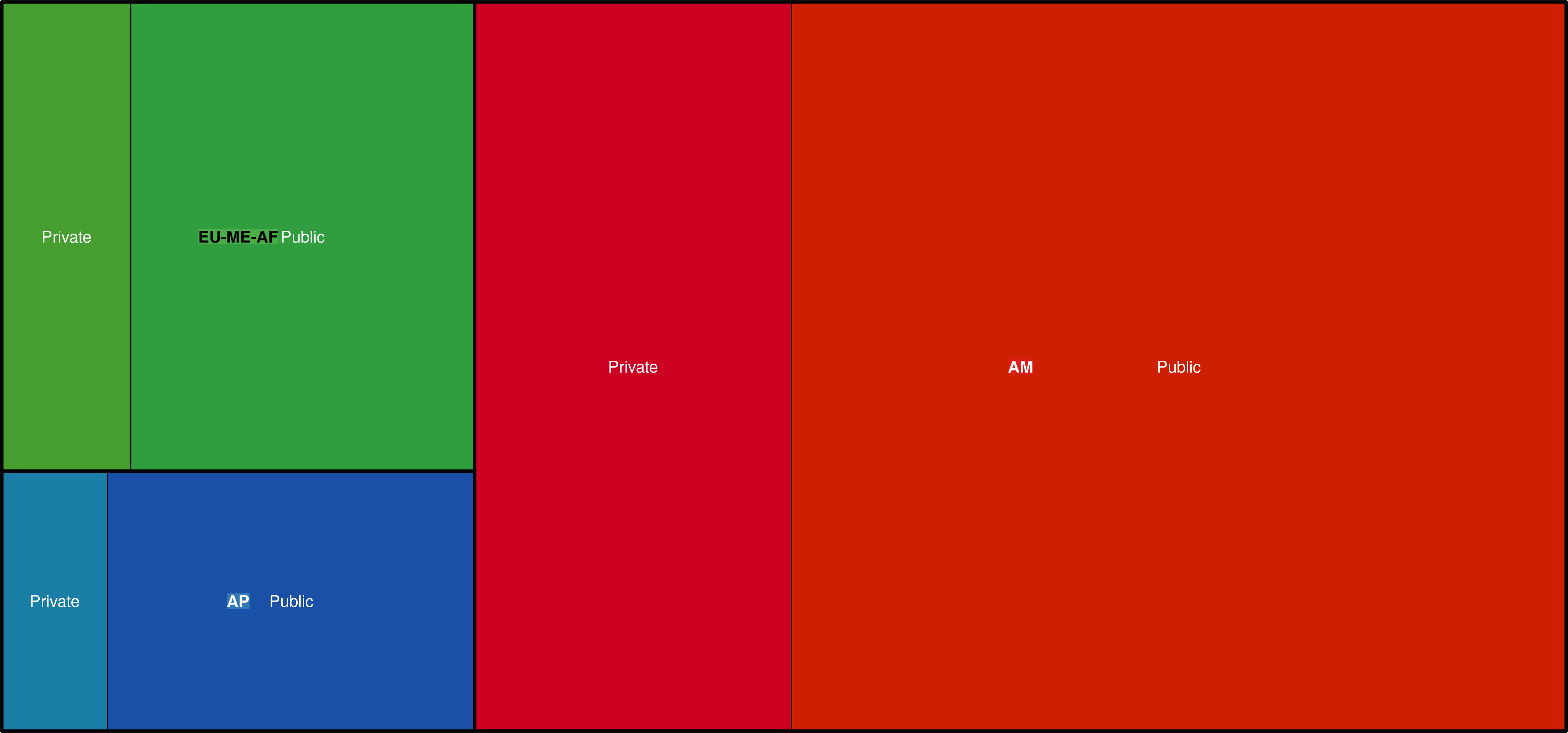}
\caption{A treemap visualization of the random stratified sample of Master in Marketing Programs AACSB-Accredited Schools.}
\label{FB}
\end{figure}

To better understand the global distribution of AACSB-accredited Master's programs in Marketing, we examine the segmentation of these programs by region. The following visualization provides insight into the geographic spread and the density of these programs worldwide, illustrating how different areas contribute to the global network of marketing education. Figure \ref{FB} shows the distribution of sampled programs by type of institutions (i.e., public versus private) and their location in the following three continental regions: the Americas (AMES), Asia-Pacific (AS-PA), and Europe-Middle East-Africa (EU-ME-AF).\\

Most of the worldwide academic offerings for Master in Marketing programs are located in the AMES region (69.77\%), followed by EU-ME-AF (19.29\%) and the AS-PA region (10.93\%). Most of these academic offering comes from public higher education institutions (71.22\%), with fewer options in private institutions (28.78\%). In the AMES region, the public-private ratio is around 2.31, lower than that of the EU-ME-AF region (2.64) and the AS-PA region (3.53). We did not collect data about tuition fees for these programs. We regard these three regions as ``sub-systems'' of an extensive system operating at a global scale by focusing on marketing education. To analyze these systems properly, it is fundamental to understand the size of the bipartite network in each of these regions. In this regard, the estimation of network size serves as a frame of reference to understand the centrality of each soft skill in one geographic location at a time, preventing misleading descriptions from cross-regional comparisons that ignore the size of these sub-systems.

\subsection{Data}

The data we collected was guided by the official public information provided by the AACSB website. According to AACSB, by the end of December 2023, 622 accredited institutions had met rigorous standards and were committed to upholding and advancing the quality of their business-related programs. We relied on the list of these higher education institutions to conduct a Google-based search to identify the explicit presence of ``Master in marketing'' programs offered by these higher education institutions. 

We visited all university websites with a Master's in marketing and downloaded its corresponding brochure as a standard portable document format (PDF) file. Most brochures were available in English and the local language of the associated country of origin. We created an open GitHub repository that stores these files as the computational data input for our analytical strategy. We identified 222 institutions with at least one explicit ``Master in Marketing'' program. We found 263 brochures offered by 190 higher education institutions out of 222, representing 85.58\% of the academic offerings in Master's in Marketing from all AACSB-accredited institutions. We discarded all brochures unavailable in English and those explicitly forbidden to be downloaded from its official website.

Data pre-processing and analysis were conducted in the \textsf{R} system \cite{RCore2022} with standard data science techniques \cite{Wickham2019} combined with text-mining procedures \cite{Benoit2018} and network-based analysis and modeling \cite{Wang2013,Lusher2013}. In using the \textsf{R} package \texttt{quanteda} \cite{Benoit2018}, we parsed 263 files to create a corpus that served as input for a standard document-term matrix. We regard this matrix as a ``skill-brochure matrix,'' where skills are arranged as rows and brochures as columns. In this matrix, cell entries can only have one numeric value out of the two options: If a word or phrase is explicitly present in a brochure, the cell entry equals one, and otherwise, it equals zero. Here, cells with non-zero values represent a link or connection between the skill and the brochure. We used the resulting rectangular matrix as the bi-adjacency matrix that represents the term $E$ in the triple $Y = (R, C, E)$ as $E \subseteq R \times C$ \cite{Kevork2022}.  

Developing the skill-brochure matrix relied on a \textit{keyword-in-context filter} \cite{Benoit2018}. This filter facilitated the search guided by words or phrases provided by the skills list associated with marketing managers as captured by \href{https://www.O*NETonline.org/link/table/details/sk/11-2021.00/Skills_11-2021-00.xlsx?fmt=xlsx&s=IM&t=-10}{O*NET}. The O*NET skills list captures a total of 35 specific skills, from \textit{Worker Requirements: Skills} section of O*NET. These skills represent the core competencies essential for successful job performance in various occupations, including the role of Marketing Manager. The Worker Requirements Skills list includes abilities directly linked to effective task execution and job success. Evaluating the alignment between academic offerings and job market demands is critical. A preliminary analysis of this list revealed that 28 of these skills are of non-null importance in the standardized scores of O*NET methodology. Figure \ref{FA} shows the Skills' Importance reported by the O*NET profile of Marketing Managers.

As marketing managers have a standard profile in all regions of the world (see Figure \ref{FA}), we used this skill list to develop a skills dictionary that served as the first partition of our bipartite network (see supplementary material). The brochures downloaded from sampled institutions serve as data input to build the second partition of our bipartite network. For example,  a program brochure with the following textual description: ``\textit{Our graduates will be able to supervise professional marketing teams to launch marketing campaigns on time, under competitive budget settings}'' will be automatically linked to skills such as management of personnel resources, time management, and management of financial resources, as indicated by O*NET's skills list for marketing managers (see the list \href{https://www.O*NETonline.org/link/summary/11-2021.00}{\textcolor{blue}{here}}). The automatic connection between brochures and skills relied on the skill dictionary we created as a list and structured data frame, using \texttt{quanteda} library \cite{Benoit2018}, following the method described by \cite{Garciachitiva2024}.  

We used the O*NET importance as a quantitative nodal attribute for the first partition in our network. This O*NET importance is an independent covariate that captures the assigned relevance to each skill as described by O*NET methodology  \cite{Peterson2001}. This O*NET importance is an empirical criterion to which we can estimate the alignment between the academic offering and the job market, provided it correlates strongly with the skills network degree estimated from our bipartite network. To complement our network with other covariates, we used the region and the school type as qualitative attributes at the nodal level in the second partition. These attributes serve as covariates to test the skill's popularity as a mechanism for node connections in our bipartite network.  

\section{The Bipartite Network}

\begin{figure}
\centering
\includegraphics[width=.7\columnwidth]{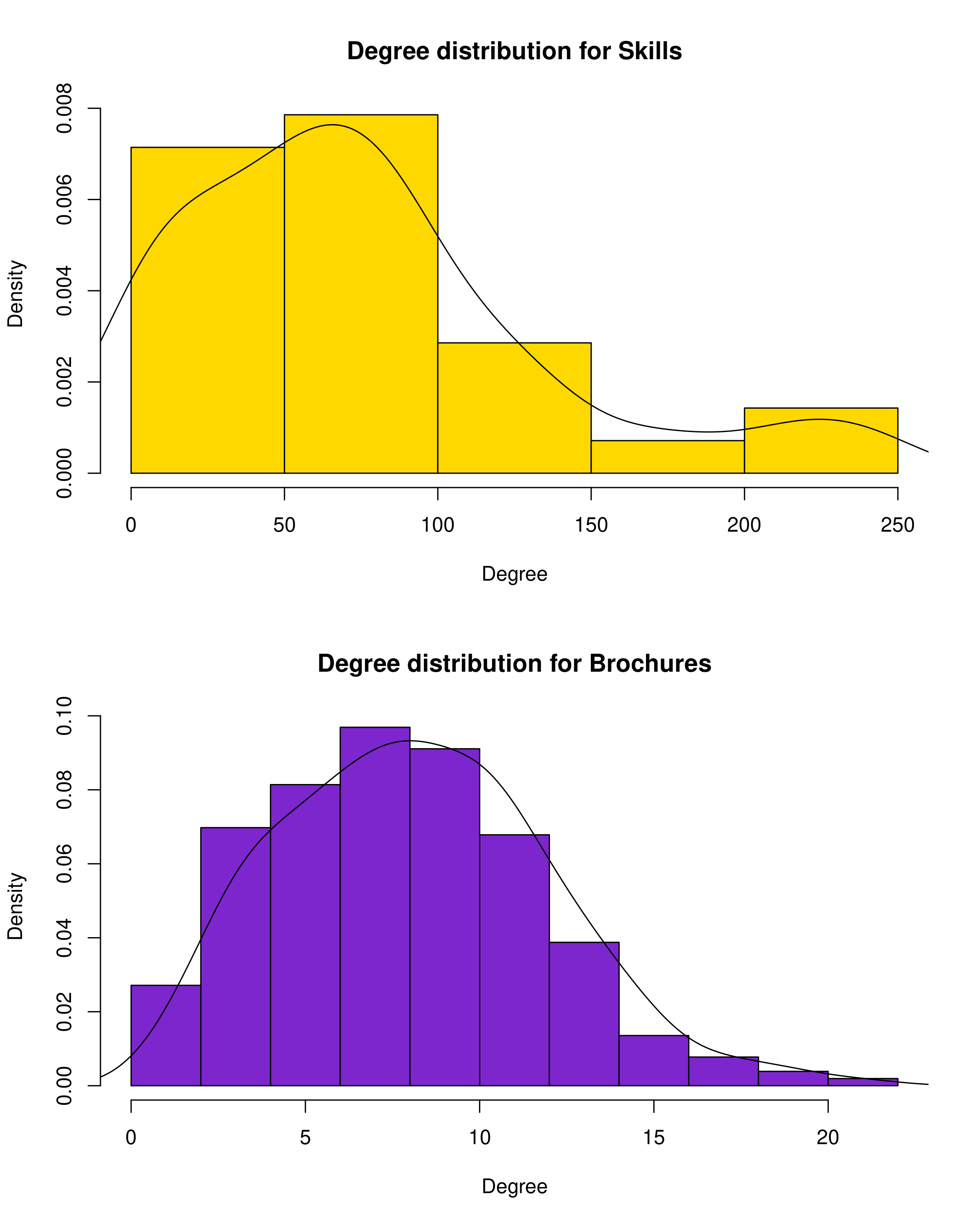}
\caption{Degree distributions for skills and brochures.}
\label{FC}
\end{figure}

The resulting bipartite network with binary entries (0/1) has 2,110 connections between 28 skills and 258 brochures from 190 institutions in three continental regions. Figure \ref{FC} shows the degree distributions for both partitions. The average degree and standard deviation of skills significantly outnumber those of brochures (i.e., on average, skills connect with 75 brochures, and brochures connect with 8 skills). This connectivity pattern for skills aligns with our expectations as they work as learning outcomes for the academic offering, so they are present in the brochures that promote their university programs.

Table \ref{T0} summarizes the number of program brochures (PB) sampled by sub-graph as well as the descriptive statistics for the degree centrality of skills and brochures. All sub-graphs show that the skills' average and standard deviation network degrees outnumber those of brochures. A curious detail occurs in the AS-PA sub-graph where the academic offering is the smallest compared with other regions in terms of program brochures sampled. In the AS-PA sub-graph the average degree for skills is the closest to the average degree of program brochures compared with the rest of the regions, pointing an incipient academic offering. 

\begin{table}
\centering
\caption{Degree centrality by sub-graphs}
\label{T0}
\begin{tabular}{lccccc}
\hline
\multirow{2}{*}{Sub-graph} & \multicolumn{3}{c}{~~~Skills} & \multicolumn{2}{c}{Brochures} \\
                        & PB  & Mean & SD &  Mean & SD           \\ \hline
Public University       & 178  & 54.07 & 42.21 & 8.20 & 3.65 \\
Private University      & 80   & 23.21 & 18.34 & 8.12 & 4.32\\
EU-ME-AF                & 126   & 42.44 & 31.67 & 9.10 & 3.99 \\
AMES                    & 100  & 26.96 & 24.25 & 7.55 & 3.41 \\
AS-PA                   & 32 & 8.04 & 6.84 & 6.53 & 3.83 \\ \hline
Entire network          & 258       & 75.36 & 60.18 & 8.18 & 3.86  \\ \hline
\end{tabular}
\end{table}

Table \ref{T1} reveals skills ranking by their ranking position in O*NET and their degree centrality in our bipartite network. It also shows each skill's degree centrality as the percentage of the total connections. We note that seven skills accumulate nearly 52\% of connections. 

\begin{table}
\centering
\caption{Skills degree centrality}
\label{T1}
\scriptsize
\begin{tabular}{lcccc} \hline
Skills                      &\multicolumn{2}{c}{Ranking by}    \\
                            & O*NET & Centrality & Degree & Perc   \\ \hline
Management of personnel resources & 5  & 1 & 239    & 11.33 \% \\
Judgment and decision making      & 2 & 2 & 214    & 10.14 \% \\
Coordination                      & 4 & 3 & 173    & 8.20 \%  \\
Active listening                  & 1 & 4 & 133    & 6.30 \%  \\
Management of financial resources & 9 & 5 & 129    & 6.11 \%  \\
Instructing                       & 8 & 6 & 113    & 5.36 \%  \\
Time management                   & 4 & 7 & 103    & 4.88 \%  \\
Mathematics                       & 10 & 8 & 92     & 4.36 \%  \\
Programming                       & 12 & 9 & 90     & 4.27 \%  \\
Active learning                   & 1 & 10 & 83     & 3.93 \%  \\
Learning strategies               & 7 & 11 & 78     & 3.70 \%  \\
Quality control analysis          & 12 & 12 & 75     & 3.55 \% \\
Management of material resources  & 11 & 12 & 75     & 3.55 \% \\
Monitoring                        & 2 & 14 & 71     & 3.36 \%  \\
Systems evaluation                & 4 & 15 & 69     & 3.27 \%  \\
Science                           & 13 & 16 & 55     & 2.61 \% \\
Speaking                          & 1 & 17 & 53     & 2.51 \%  \\
Reading comprehension             & 1 & 18 & 51     & 2.42 \%  \\
Critical thinking                 & 1 & 19 & 50     & 2.37 \%  \\
Complex problem solving           & 3 & 20 & 41     & 1.94 \%  \\
Service orientation               & 7 & 21 & 37     & 1.75 \%  \\
Persuasion                        & 2 & 22 & 31     & 1.47 \%  \\
Technology design                 & 13 & 23 & 17     & 0.81 \% \\
Negotiation                       & 3 & 24 & 13     & 0.62 \%  \\
Writing                           & 6 & 25 & 13     & 0.62 \%  \\
Operations monitoring             & 13 & 26 & 6      & 0.28 \% \\
Social perceptiveness             & 1 & 27 & 4      & 0.19 \%  \\
Operations analysis               & 5 & 28 & 2      & 0.09 \%  \\ \hline
Total of connections              &   &    & 2110   & 100\%  \\   \hline
\end{tabular}
\end{table}

\begin{figure}
\centering
\includegraphics[width=.7\columnwidth]{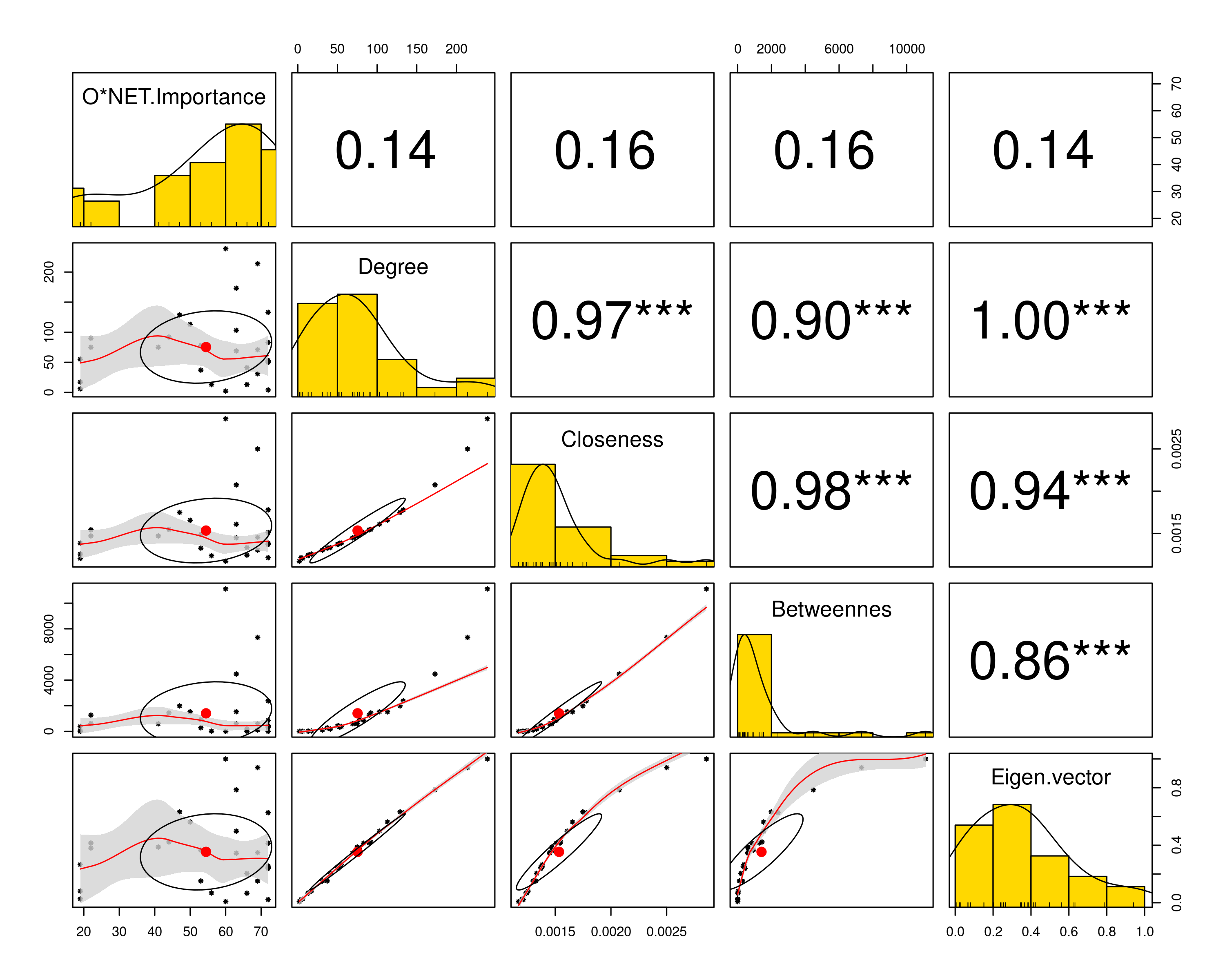}
\caption{Correlation plot for skills' importance captured by O*NET and our network-based centrality metrics.}
\label{FD}
\end{figure}

Table \ref{T1} shows that the ranking for these skills does not match when comparing their position by O*NET and their degree centrality in our bipartite network, which suggests a qualitative misalignment between the job demands for marketing managers as captured by O*NET and the academic offering as captured by their derived degree centrality from our bipartite network. Figure \ref{FD} shows this mismatch as the correlations between O*NET-based skill's importance with all of our network-based centrality metrics are low and non-significant. In contrast, the correlations among our estimated metrics are all high and statistically significant. This correlation pattern provides evidence about the potentially weak and non-systematic alignment between job requirements for marketing managers as captured by O*NET and the AACSB-related academic offering. With this network data description, we now switch the attention to the illustration of our research design and model specifications, including network-based statistics and network covariates.

\subsection{Research design and Models specification}

We rely on the bipartite exponential random graph model (ERGM) as a research design framework \cite{Wang2009,Wang2013}. Despite early proposals on bipartite network analysis \cite{Borgatti1997,Asratian1998,Skvoretz1999}, the development of ERGM as a statistical inferential tool is more recent \cite{Lusher2013,Kevork2022,Wang2024}. 

As per \cite{Kevork2022}, the probability of observing a bipartite network $Y$ depends on a set of network statistics as follows:

\begin{equation}
P(Y = y|\theta,u,v) = \frac{exp\left\{\theta^T s(y) + u^T r(y) + v^T c(y)\right\}}{\kappa(\theta, u, v)}
\end{equation}

where $\theta \in \mathbb{R}^q$ is a $q$ dimensional parameter vector, $s(y) = (s_0(y), s_1(y), ...)$ is a $q$ dimensional vector of network statistics with $s_0(y) = \sum_i\sum_j y_{i,j}$ and $r(y) = (\sum_{j=1}^{m} y_{1j},\sum_{j=1}^{m} y_{2j}...)$ is the $n$ dimensional vector of node degrees in $R$ and $c(y) = (\sum_{i=1}^{n} y_{i1},\sum_{i=1}^{n} y_{i2},...)$ is the $m$ dimensional vector of node degrees in $C$. The $\kappa(\theta, u, v)$ is a normalizing factor that ensures that $\mathbb{P}(Y = y|\theta, u, v)$ is a legitimate probability mass function. The vector $s(y)$ includes any structural configurations of the bipartite network, which can be extended to $s(y, X)$ should covariates $X$ are included in the model, see \cite{Kevork2022} for additional details. 

From this viewpoint, the distinction between endogenous and exogenous mechanisms that may affect the connection between skills and brochures enables the network researcher to specify a variety of bipartite degree statistics. Endogenous statistics refer to specific local network configurations in the observed network as coined by \cite{Wang2009,Wang2013}. Given that our network data shows that, on average, skills connections outnumber program brochures at the nodal level, we have sufficient evidence to assume that popular skills will play a significant role in predicting network ties with positive signs in their ERGM estimates. The selection of the most popular skills is straightforward from Table \ref{T1}. If our assumption is correct, including the skills that occupy the top five positions in Table \ref{T1} should suffice to achieve statistical significance in an exponential random graph model. The specifications of Model 1 and Model 2 capture these assumptions for the top 3 and top 5 most popular skills, respectively.

As skill's popularity is intrinsically related to their explicit presence in program brochures, we also have sufficient evidence to pose that homophily-based effects will play a significant role in predicting network ties with positive signs in their ERGM estimates.  Roughly speaking, homophily refers to the idea that network ties are more likely to form between nodes who share one or more attributes in common than would otherwise be expected. The specifications of Model 3 and 4 capture this assumption over the specification of Model 1 and 2. Following \cite{Kevork2022}, homophilic effects can be estimated by summing $y_{ik} y_{jk}$ for all nodes in our first partition (i.e., skills) of the same type $i$ and $j$ and all program brochures $k$, with the network statistics \texttt{nodematch} as follows:

\begin{equation}
nodematch(y) = \frac{1}{2}\sum_{i = 1}^{n}\sum_{j = 1}^{n}\sum_{k = 1}^{m} y_{ik}y_{jk}y_{jk}I\left\{ i~matches~j,~i\neq j \right\}
\end{equation}

Given the empirical setting of our bipartite network, we can also use three specific covariates that will help us to test this homophilic behavior for network ties. The degree centrality by sub-graphs summarized in Table \ref{T0} shows that skills' degrees in brochures from public universities are higher than their private counterparts. This pattern suggests that program brochures offered by public universities can play a statistical significant role in predicting network ties. Likewise, skills' degree in the central hemisphere of the world (EU-ME-AF) is higher than the Western (AMES) and the Asian-Pacific (AS-PA) regions, suggesting that program brochures offered by universities in the European-Middle East-African region can play a significant role in predicting network ties. If our assumptions are correct, the ERGM estimates for these covariates will be positive and statistically significant. A final assumption from our bipartite network relates to the empirical alignment between the Skills' importance reported by the O*NET profile of Marketing managers and the centrality of the same skills captured in our bipartite network. Figure \ref{FD} showed evidence of a weak and non-systematic alignment. If our assumption is correct, we should confirm this lack of alignment with non-significant, near-zero ERGM estimates. Based on all these arguments, we now switch the attention to our model specifications.

\textbf{Model 1}. This initial model set the individual effect of the top-3 most popular skills from Table 2 (i.e., management of personnel resources, judgment and decision making, and coordination) on network ties. Estimating these effects is compared with the effect of \textit{edges}, which is the baseline propensity of skills appearing in program brochures (i.e., university authorities and faculty recognize that students enrolling in these programs will develop these skills). Thus, the statistical estimate for \textit{edges} is the benchmark to determine whether its influence is positive or negative and statistically significant or not on network ties. The fundamental hypothesis under scrutiny in this model is that skill's popularity is a valid mechanism that accounts for node connections in our network. From this model, we hypothesize that statistical estimates will be highly significant with positive values.  

\textbf{Model 2}. This model extends the previous model by considering the top-5 most popular skills from Table \ref{T1}, adding to the first three, the skills of active listening and management of financial resources. The comparison between Model 1 and Model 2 allows us to estimate the hypothesis of skill's popularity can be extended to additional competencies. The fundamental hypothesis to test with this model is that the skill's popularity is a valid mechanism that accounts for node connections in our network. We expect that all statistical estimates will be highly significant with positive values.  

\textbf{Model 3}. Unlike the first two models, Model 3 includes a network statistic that captures the homophilic behavior of skill's importance captured by O*NET and sets this importance as an exogenous factor for nodes in the first partition of our bipartite network. The fundamental hypothesis in this model is that skill's popularity for the top 3 skills coexist with the homophilic behavior of skills with a systematic presence in program brochures, regardless their rank. Our hypothesis is that statistical estimates will be highly significant with positive values. 

\textbf{Model 4}. This model extends the scope of Model 3 by  probing the mechanism of skills popularity for the top-5 most popular skills and the homophily mechanism of skills' importance captured by O*NET and modeled as an exogenous factor for nodes in the first partition of our bipartite network. Like previous models, we hypothesize that statistical estimates will be highly significant with positive values.  

\textbf{Model 5}. This model proposes the hypothesis that skills' popularity and homophily coexist as statistical significant mechanisms that explain ties in our bipartite network. In addition to these mechanisms, we hypothesize that the connections between skills and programs are more likely to occur in  in public universities from the European-Middle East-African region. Our expectation is that statistical estimates will be highly significant with positive values.

\textbf{Model 6.} This is our final model and as such it has the most complex model specification compared with the rest. Here, our fundamental claim is that skills' popularity can be extended, once again, from the top-3 to the top-5 most popular skills. With such an extension, the statistical estimates for all of these skills will be highly significant with positive values in decreasing order. In addition, skills' homophily will coexist with positive ERGM estimates associated with program brochures from public universities and the EU-ME-AF region. Like the rest of our models, we expect that all statistical estimates will be highly significant with positive values.

We specified and ran all our models in the \textsf{R} system \cite{RCore2022} with the help of the following libraries: \texttt{igraph} \cite{Csardi2006}, \texttt{network} \cite{Butts2008}, and \texttt{ergm} \cite{Krivistky2023}. As usual in the analysis of network data, we followed the simulation-based strategy to assess goodness of fit indices for reporting parameter estimates \cite{Lusher2013,Cranmer2021}. In doing so, for each estimated model, we sampled 1000 networks preserving the bipartite network structure of models in Table \ref{T2}. Our supplementary material provides additional computational details of these models for those interested in replicating our results, following standards of reproducible research \cite{Gandrud2018}.

\section{Results}

We begin our results by summarizing the ERGM estimates for our six Models. The main hypothesis under scrutiny for Models 1 and 2 is that of skills' popularity. Models 3 and 4 extend this hypothesis to the existence of homophily among these skills. Finally, Models 5 and 6 propose that skills' popularity and homophily coexist with exogenous factors that explain network ties. We proposed that these factors are related to the academic offering of public universities from the European-Middle East-Africa region.

\begin{table}
\centering
\caption{ERGM estimates (standard error in parentheses)}
\label{T2}
\scriptsize
\begin{tabular}{lcc}\hline
\textbf{Effect}    & \textbf{Model 1} & \textbf{Model 2}  \\
Density & -1.207(0.029)*** & -1.349(0.032)***    \\
Management of personnel resources & 3.739(0.240)***  & 3.881(0.240)***   \\
Judgment and decision making & 2.789(0.168)*** & 2.931(0.168)***  \\
Coordination & 1.918(0.135)*** & 2.060(0.136)***  \\
Active listening &  & 1.411(0.128)***  \\
Management of financial resources & & 1.349(0.128)***  \\  \hline
\hfill AIC & 7664 & 7461  \\
\hfill BIC & 7692 & 7502  \\
\hfill Mahalanobis distance & 0.343 & 0.327  \\ \hline
\textbf{Effect}  & \textbf{Model 3} & \textbf{Model 4}  \\
Density & -1.292(0.039)*** & -1.440(0.042)***   \\
Management of personnel resources & 3.804(0.241)*** & 3.943(0.244)*** \\
Judgment and decision making & 2.845(0.168)*** &  2.990(0.170)*** \\
Coordination & 1.941(0.137)*** & 2.091(0.137)***  \\
Active listening &  & 1.337(0.132)***  \\
Management of financial resources & & 1.439(0.131)***     \\  
Nodematch(O*NET Importance) & 0.141(0.041)***  & 0.145(0.041)*** \\ \hline
\hfill AIC & 7656 & 7452  \\
\hfill BIC & 7690 & 7500  \\
\hfill Mahalanobis distance & 0.163 &   \\ \hline
\textbf{Effect}  & \textbf{Model 5} & \textbf{Model 6}  \\
Density & -1.215(0.059)***  & -1.359(0.062)***    \\
Management of personnel resources & 3.800(0.236)*** & 3.945(0.238)***\\
Judgment and decision making & 2.848(0.169)*** & 2.997(0.171)***\\
Coordination & 1.941(0.134)*** & 2.094(0.137)***\\
Active listening &  & 1.343(0.131)***\\
Management of financial resources & &1.445(0.132)***  \\  
Public University & -0.098(0.060) & -0.101(0.061)$^\bullet$ \\ 
EU-ME-AF & -0.021(0.056) & -0.02(0.057) \\ 
Nodematch(O*NET Importance) & 0.140(0.041)*** &  0.144(0.041)***\\ \hline
\hfill AIC & 7657  & 7452  \\
\hfill BIC & 7705 & 7514  \\
\hfill Mahalanobis distance & 0.068 & 0.158  \\ \hline

$^{\bullet } p <$ 0.1; * $p <$ 0.05; **  $p <$ 0.01; *** $p <$ 0.001
\end{tabular}
\end{table}
\newpage

Table \ref{T2} shows that the density effect is negative and highly significant for all our model specifications. This finding pinpoints that the connectivity between skills and programs is quite below any expectation from chance, suggesting that the academic offering focused on skills is in an early stage. Even though this evidence goes against our expectations, it serves as the benchmark to which we can compare other ERGM effects. Our first hypothesis proposed a positive and highly significant estimate for skills' popularity as an endogenous factor that accounts for network ties. We confirmed this hypothesis by looking at the ERGM estimates of Model 1 and 2. Our second hypothesis was that skills have an homophilic behavior in terms of their alignment with the job demands captured by the O*NET importance scores for skills of marketing managers. We confirmed this alignment with Model 3 and 4 where the estimates showed the highly significant weakest values. Our final hypothesis was that skills' popularity and homophily coexist with the influential role of public universities from the European-Middle East-Africa region.  This hypothesis is not true for Model 5 and 6 where the ERGM estimates proved to be negative and non-significant at the traditional level of $\alpha \leq $ 0.01. Table \ref{T4} summarizes the goodness-of-fit statistics for these models by comparing the observed statistics of our bipartite network (Obs) with its simulated statistics (i.e., Min, Mean, Max, p) in all six models.

\begin{table}
\centering
\caption{Goodness-of-ﬁt statistics for Models 1–6.}
\label{T4}
\scriptsize
\begin{tabular}{lcccccccccc}
\hline
\multirow{2}{*}{\begin{tabular}[c]{@{}l@{}}Estimated\\ parameter\end{tabular}} & \multicolumn{5}{c}{\textbf{Model 1}} & \multicolumn{5}{c}{\textbf{Model 2}} \\ \cline{2-11}
& Obs  & Min & Mean & Max & p & Obs  & Min & Mean & Max & p \\ \hline
Density & 2110 & 2019 & 2105.91 & 2181 & 0.96 & 2110 &2057 &2109.90 &2200 &0.98   \\
Management of personnel resources & 239 & 228 & 239.58 & 250 & 1.00 &239 &230 & 239.23& 248&0.98   \\
Judgment and decision making & 214 & 196 & 212.98 & 228 & 0.90 & 214 & 199 & 215.05 & 226 & 0.90   \\
Coordination & 173 & 150 & 173.44 & 192 & 1.00 & 173
& 153 &172.40 & 190&0.98   \\
Active listening & & & & & & 133& 113 & 132.55& 155&0.98   \\
Management of financial resources & & & & & & 129& 103& 127.91&148 &1.00 \\ \hline
\multirow{2}{*}{\begin{tabular}[c]{@{}l@{}}Estimated\\ parameter\end{tabular}} & \multicolumn{5}{c}{\textbf{Model 3}} & \multicolumn{5}{c}{\textbf{Model 4}} \\ \cline{2-11}
& Obs  & Min & Mean & Max & p & Obs  & Min & Mean & Max & p \\ \hline
Density & 2110 & 2026 & 2109.81 & 2160 & 1.00 &2110  &2036 &2107.97 &2183 &0.90   \\
Management of personnel resources & 239 & 230 & 239.25 & 250&1.00 &239 &228 &239.64 &249 &0.98   \\
Judgment and decision making & 214 & 197 & 214.60 & 229& 0.98&214 &196 &214.87 &227 &0.88   \\
Coordination & 173 & 153 & 172.17& 186& 0.96&173 &148 & 172.96& 192&0.98   \\
Active listening & & & & & &133 &109 &132.91 &152 &1.00   \\
Management of financial resources & & & & & &129 &113 &128.30 &145 &0.90   \\ 
Nodematch(O*NET) & 582 & 521 & 584.83&667 & 0.80&582 &516 &584.95 & 652&0.94   \\ \hline
\multirow{2}{*}{\begin{tabular}[c]{@{}l@{}}Estimated\\ parameter\end{tabular}} & \multicolumn{5}{c}{\textbf{Model 5}} & \multicolumn{5}{c}{\textbf{Model 6}} \\ \cline{2-11}
& Obs & Min & Mean & Max & p & Obs  & Min & Mean & Max & p \\ \hline
Density & 2110 & 2011 & 2110.96 & 2222 &0.98  & 2110&1999&2115.37& 2198&0.90   \\
Management of personnel resources & 239 & 231 & 239.42 & 251 &1.00&239 &229 &238.64 &248 &0.98   \\
Judgment and decision making &214&191&214.45&229&1.00 &214 &199 & 213.80&229 &0.98   \\
Coordination &173&152&174.38&193&0.92&173&147&172.93&188&1.00   \\
Active listening & & & & & &133&115&133.06&155&0.94   \\
Management of financial resources & & & & & &129&108&129.36&151&1.00   \\
Nodematch(O*NET) &582 &492 &581.05 &665 &0.96 &582 &488 &587.56 &675 &0.92   \\
Public University &1426 &1363 &1424.52 &1503 &0.94 &1426 &1343 &1428.71 &1500 &1.00  \\
EU-ME-AF &1017 & 959&1014.77 &1081 &0.98 & 1017&953 &1022.95 &1082 &0.80 \\ \hline
\end{tabular}
\end{table}

\newpage
\section{Discussion and conclusions}

The American economy is gliding to skills-based hiring, which favors demonstrated skills over university degree completion \cite{Fuller2022}. Thanks to this policy, some efforts focus on understanding the importance of skills that, for example, marketing managers should develop for new product development and sustainable business performance \cite{Ali2020}. Other efforts include the implications for the academic offering and its relevance for the 21st-century workforce \cite{Vista2020} to identify skill discrepancies between research, education, and jobs \cite{Borner2018} and evaluate education-occupation match \cite{Gadar2018}. On top of these efforts, recent research shows compelling evidence for using bipartite network modeling to understand the relationship between professional jobs, skills, and education \cite{Aufiero2024, Garciachitiva2024}. Our contribution leverages bipartite exponential random graph modeling \cite{Kevork2022} as a framework that facilitates the analysis of the alignment between marketing managers' professional skills and AACSB-accredited graduate programs of marketing. As a case study, our work provides relevant theoretical, methodological, and practical contributions.

Our first contribution is theoretical, reinforcing the foundations of a new approach known as ``network science in education'' aiming to develop methods, curricular materials, and resources for learning and teaching in higher education \cite{Cramer2018}. From this approach, we claim that future education will have a non-trivial combination of skills-based training with flexible and adaptable contents that fulfill the shifting demands of the labor market. The uniqueness of this education won't be a static goal but an ever-changing objective that facilitates how academic institutions can better serve students and industry needs. Although this contribution is consistent with broader trends in educational policy that emphasize the necessity for flexible curricula capable of responding to market demands, as highlighted in studies on higher education adaptability \cite{Schlegelmilch2020}, our work is the first of its kind in developing an empirical effort based on bipartite network modeling that did not depend on traditional data based on surveys or questionnaires, but on the academic offering available and analyzable through text mining techniques. By mapping the skills and university programs as a bipartite network structure, we showed that its connectivity is significantly lower than expected by chance revealing negative and statistically significant density effects, which is an indicator that skills-based education is in its early stage of development.  

Our second contribution is methodological. Our case study focused on marketing graduate programs to show that bipartite network modeling is an effective tool that facilitates the estimation of the empirical alignment between the academic offering of graduate programs and industry demand for marketing managers. The extension of this analysis is straightforward to any other occupation. The realization of this analysis takes professional profiles in demand by the job market (e.g., O*NET) and set them as exogenous factors that may affect the connection between skills and programs in specific occupations. Thus, our work is the first of its kind to provide fresh evidence that illustrates the utility of bipartite exponential random graph modeling \cite{Wang2009,Wang2013,Kevork2022} in exploring hypothetical mechanisms that account for skills-programs connections (e.g., homophily, popularity, alignment). Our case study also showed that the statistical estimation of these mechanisms has practical implications for the design and evaluation of graduate marketing programs, as they can help ensure that these programs match the skills demanded by the job market. 

As a result of the contributions mentioned above, our case study provides practical takeaways. While descriptive analyses suggest a qualitative lack of alignment between the skills' ranking by job demands captured by O*NET and our bipartite network, inferential analyses based on exponential random graph model estimates showed that skills' popularity and homophily coexist with a weak yet systematic alignment to marketing mangers' job-related skills. The emphasis on ``Management of Personnel Resources,''  ``Judgment and Decision Making,'' ``Coordination,'' ``Active listening,'' and ``Management of financial resources'' underlines their prominence in academic brochures, challenging prevailing trends in marketing education that increasingly emphasize managerial and decision-making skills, often at the expense of communication and critical thinking abilities. From this perspective, we concur with \cite{Schlegelmilch2020} that these skills should not belittle core communication abilities such as speaking, writing, and understanding complex information, which are critical for effective leadership and stakeholder management. This focus is crucial for enhancing the employability of graduates and ensuring that educational offerings remain relevant in an ever-evolving job market \cite{martinez2022,stevens2021}. We suggest that closing the gap between the high-value skills emphasized by O*NET and their explicit mention in marketing program brochures is crucial for aligning academic offerings with the actual demands of the marketing industry. Unlike more targeted forms of marketing education (e.g., executive programs, certifications, and other emerging modalities that focus on specific skills), comprehensive marketing programs should emphasize not only the relevance of these competencies to employers but also their potential as key differentiators for individuals aspiring to leadership roles in the field \cite{Anderson2023,williams2023}.

The focus on students and industry needs is a call for faculty members to increase their cooperation with non-academic stakeholders. Incorporating real-time industry feedback into curriculum development and promotional strategies is essential for keeping programs at the forefront of industry developments and technological progress. This proactive stance, driven by industry input, can also help prevent curricula from becoming outdated—a common challenge in fast-evolving fields like marketing \cite{martinez2022}. This connection is critical in today's competitive academic environment, where the ability to attract and retain students depends mainly on how valuable and relevant the program's offerings are perceived \cite{Bennett2009}. Additionally, aligning promotional content with industry needs can enhance the institution's reputation and brand equity, making these programs more appealing to students and employers who seek graduates ready to contribute from day one \cite{stevens2021}.

Our case study is not free of limitations. Previous studies of network science in education have been focused on mapping the curricular structure and contents by analyzing the syllabi or course schedules  \cite{Sayama2018}. Such a focus for research purposes might be prohibitive given that syllabi are not publicly available in online multimedia summaries of marketing graduate programs. Nonetheless, syllabi can be used as data input for reproducing our analysis in professional benchmarking analysis conducted by members of the AACSB community. Likewise, further studies can extend the analysis to other graduate programs above and beyond marketing and AACSB accreditation standards.

\section*{CRediT Author's contribution}

\textbf{Conceptualization}: MPG-C and JCC; \textbf{Data Curation}: JCC; \textbf{Formal Analysis}: JCC; \textbf{Investigation}: MPG-C, SD, and JCC; \textbf{Methodology}: MPG-C and JCC; \textbf{Project Administration}: JCC; \textbf{Resources}: MPG-C, SD, and JCC; \textbf{Supervision}: SD and JCC; \textbf{Validation}: JCC; \textbf{Visualization}: JCC; \textbf{Writing – Original Draft Preparation}: SD and JCC; \textbf{Writing – Review \& Editing}: JCC  

\bibliographystyle{abbrv}  


\end{document}